\documentclass[]{spie}  %>>> use for US letter paper
\usepackage[]{graphicx}

\title{Metal-insulator transition and glassy behavior in two-dimensional
electron systems}

\author{Dragana Popovi\'c$^{\ast}$\supit{a}, Sne\v{z}ana
Bogdanovich$^{\dag}$\supit{a}, J. Jaroszy\'nski$^{\ddag}$\supit{a},
T. M. Klapwijk\supit{b}
\skiplinehalf
\supit{a}National High Magnetic Field Laboratory, Florida State University,
1800 E. Paul Dirac Drive,\\Tallahassee, FL 32310, USA\\
\supit{b}Department of Applied Physics, Delft University of Technology,
2628 CJ Delft, \\The Netherlands}

\authorinfo{$^{\ast}$Corresponding author. E-mail:
dragana@magnet.fsu.edu; phone 1 850 644-3913\\
\hspace*{0.22in}$^{\dag}$Present address: General Electric Corporate R\&D, KW
B322, 1 Research Circle, Niskayuna, NY 12309; E-mail:
bogdanov@crd.ge.com; phone 1 518 387-6424\\
\hspace*{0.22in}$^{\ddag}$Permanent address: Institute of Physics,
Polish Academy of Sciences, al. Lotnik\'ow 32/46, 02668 Warszawa,
Poland; E-mail: jaroszy@magnet.fsu.edu; phone 1 850 644-5699\\
T. M. K.: E-mail:
klapwijk@dimes.tudelft.nl; phone 31 15 278-2600}
\begin{document}
  \maketitle

%%%%%%%%%%%%%%%%%%%%%%%%%%%%%%%%%%%%%%%%%%%%%%%%%%%%%%%%%%%%%
\begin{abstract}
Studies of low-frequency resistance noise demonstrate that glassy
freezing occurs in a two-dimensional electron system in silicon in
the vicinity of the metal-insulator transition (MIT).  The width
of the metallic glass phase, which separates the 2D metal and the
(glassy) insulator, depends strongly on disorder, becoming
extremely small in high-mobility (low-disorder) samples.  The
glass transition is manifested by a sudden and dramatic slowing
down of the electron dynamics, and by a very abrupt change to the
sort of statistics characteristic of complicated multistate
systems.  In particular, the behavior of the second spectrum, an
important fourth-order noise statistic, indicates the presence of
long-range correlations between fluctuators in the glassy phase,
consistent with the hierarchical picture of glassy dynamics.
\end{abstract}

%>>>> Include a list of keywords after the abstract

\keywords{metal-insulator transition, glass transition, noise,
two-dimensional systems}

%%%%%%%%%%%%%%%%%%%%%%%%%%%%%%%%%%%%%%%%%%%%%%%%%%%%%%%%%%%%%
\section{INTRODUCTION}
\label{sect:intro}

Since the development of the scaling theory of localization for
noninteracting electrons \cite{AALR}, it was widely believed that
all electronic states are localized in a two-dimensional (2D)
disordered system.  Strictly speaking, however, the possibility of
a 2D metallic state and the metal-insulator transition (MIT) in
case of strong electron-electron interactions has been an open
issue from the theoretical point of view. Recent experiments on
dilute 2D electron and hole systems in semiconductor
heterostructures \cite{SAK2001} have provided considerable
evidence for the existence of a true, zero-temperature MIT, and
revived interest in the problem of the interplay of disorder and
electron-electron interactions in 2D.  However, the physics behind
the apparent MIT is still not understood.

It is now well established that the MIT occurs in the regime where
both Coulomb (electron-electron) interactions and disorder are
strong.  Theoretically, it is well known~\cite{eglass1,eglass2}
that, in the strongly localized limit, the competition between
electron-electron interactions and disorder leads to glassy
dynamics (electron or Coulomb glass).  Some glassy properties,
such as slow relaxation phenomena, have been indeed observed in
various insulating thin
films~\cite{films21,films22,films23,films24,films25,films31,films32,films4}.
Furthermore, recent work~\cite{newgang} has suggested that the
critical behavior near the 2D MIT may be dominated by the physics
of the insulator, leading to the proposals that the 2D MIT can be
described alternatively as the melting of the Wigner \cite{Sudip},
Coulomb \cite{Thakur1,Thakur2,Pastor}, or spin glass
\cite{Sachdev1,Sachdev2}. It is clear that understanding the
nature of the insulator represents a major open issue in this
field.

While glassy systems exhibit a variety of
phenomena~\cite{glasses}, studies of metallic spin glasses have
demonstrated~\cite{Weiss93} that mesoscopic, {\it i.~e.} transport
noise measurements are required in order to provide definitive
information on the details of glassy ordering and dynamics.  We
have employed a combination of transport and low-frequency
resistance noise measurements \cite{SBPRL02,JJPRL02} to probe the
glassy behavior and the MIT in a 2D electron system (2DES) formed
in Si metal-oxide-semiconductor field-effect transistors
(MOSFETs). We find that glassy freezing occurs in the regime of
very low conductivity $\sigma$, apparently as a precursor to the
MIT. The glass transition is manifested by a sudden and dramatic
slowing down of the electron dynamics and by an abrupt change to
the sort of statistics characteristic of complicated multistate
systems.  The properties of the entire glass phase are consistent
with the hierarchical picture of glassy dynamics, similar to spin
glasses with long-range correlations. We also show that the width
of the metallic glass phase, which separates the 2D metal and the
glassy insulator, depends strongly on disorder, becoming very
small in samples with relatively low disorder.  These results are
consistent with the model that describes the 2D MIT as the melting
of a Coulomb glass \cite{Pastor,Denis,Darko}.

%%%%%%%%%%%%%%%%%%%%%%%%%%%%%%%%%%%%%%%%%%%%%%%%%%%%%%%%%%%%%
\section{SAMPLES AND EXPERIMENTAL TECHNIQUE}
\label{sect:samples}

Measurements were carried out on a 2DES in MOSFETs that were
fabricated on the (100) surface of Si. In such a device, the
disorder is due to the oxide charge scattering (scattering by
ionized impurities randomly distributed in the oxide within a few
\AA\, of the interface) and to the roughness of the Si-SiO$_{2}$
interface \cite{AFS}.  While the former dominates at low carrier densities
$n_s$,
the increasing $n_s$ improves screening and reduces the effective
disorder due to oxide charges. However, at the same time, the
electrons are pushed closer to the interface, and surface
roughness scattering becomes more important, becoming dominant at
high $n_s$.  As a result of this competition, the mobility of the
2DES exhibits a peak as a function of $n_s$.  The value of the
peak mobility at 4.2~K is usually taken as a rough measure of the
disorder \cite{AFS}. We have studied two sets of devices with the
peak mobilities that differ by about a factor of 40, which,
together with the substantial differences in their geometry, size,
and many fabrication details, means that these samples span
essentially the entire range of Si technology.

The low-mobility (\textit{i. e.} low peak mobility or high
disorder) devices were fabricated using standard 0.25~$\mu$m Si
technology \cite{Taur} with poly-Si gates, self-aligned
ion-implanted contacts, substrate doping $N_a\sim
2\times10^{17}$cm$^{-3}$, oxide charge
$N_{ox}=1.5\times10^{11}$cm$^{-2}$, and oxide thickness
$d_{ox}=50$~nm.  Their peak mobility at 4.2~K was only
0.06~m$^2$/Vs.  Most of the measurements were performed on a
1~$\mu$m long, 90~$\mu$m wide rectangular sample.  The
fluctuations of current $I$ ({\it i.~e.} $\sigma$) were measured
as a function of time $t$ in a two-probe configuration using an
ITHACO 1211 current preamplifier and a PAR124A lock-in amplifier
at $\sim 13$~Hz.  The excitation voltage $V_{exc}$ was kept
constant and low enough (typically, a few $\mu$V) to ensure that
the conduction was Ohmic.  A precision DC voltage standard (EDC
MV116J) was used to apply the gate voltage $V_g$, which controls
the carrier density $n_s$.  The current fluctuations as low as
$10^{-13}$~A were measured at $0.13\leq T\leq 0.80$~K in a
dilution refrigerator with heavily filtered wiring.  Relatively
small fluctuations of temperature $T$, $V_g$, and $V_{exc}$ were
ruled out as possible sources of the measured noise, since no
correlation was found between them and the current fluctuations.
In addition, a Hall bar sample from the same wafer was measured at
$T=0.25$~K in both two- and four-probe configurations, and it was
determined that the contact resistances and the contact noise were
negligible.

High-mobility (low-disorder) samples had the peak mobility of
$\approx 2.5$~m$^2$/Vs at 4.2~K.  They were fabricated in a Hall
bar geometry with Al gates, $N_{a}\sim10^{14}$cm$^{-3}$, and oxide
thickness $d_{ox}=147$~nm
\cite{heemskerksamples1,heemskerksamples2}. The resistance was
measured down to $T=0.24$~K using a standard four-probe ac
technique (typically $2.7$~Hz) in the Ohmic regime. The DC voltage
standard was used to apply $V_g$. Contact resistances and their
influence on noise measurements were minimized by using a
split-gate geometry, which allows one to maintain high $n_{s}$
($\approx10^{12}$~cm$^{-2}$) in the contact region while allowing
an independent control of $n_s$ of the 2D system under
investigation in the central part of the sample ($120\times
50~\mu$m$^2$) (Fig.~\ref{raverage} inset). Nevertheless, care was
taken to ensure that the observed noise did not come from either
the current contacts or the regions of gaps in the gate. For
example, since the noise measured across a resistor connected in
series with the sample and having a similar resistance was at
least three times lower than the noise from the central part of
the sample, the effect of the contact noise on the excitation
current $I_{exc}$ could be easily ruled out. Similarly, the
resistance and the noise measured between the voltage contact in
the region of high $n_s$ ({\textit e.~g.} \#5 in
Fig.~\ref{raverage} inset) and the one in the central part (\#6)
were much smaller than those measured between contacts in the
central part ({\textit e.~g.} \#6 and \#7).  In fact, they were in
agreement with what is expected based on the geometry of the
sample, which proves that the gap regions did not contribute to
either the measured resistance or noise. In order to minimize the
influence of fluctuations of both $I_{exc}$ and $T$, some of the
noise measurements were carried out with a bridge configuration
\cite{Sco87}.  The difference voltage was detected using two
PAR124A lock-in amplifiers, and a cross-spectrum measurement was
performed with an HP35665A spectrum analyzer in order to reduce
the background noise even further \cite{Verbruggen89}.  The output
filters of the lock-in amplifiers and/or spectrum analyzer served
as an antialiasing device.  Most of the noise spectra were
obtained in the $f=(10^{-4}-10^{-1})$~Hz bandwidth, where the
upper bound was set by the low frequency of $I_{exc}$, limited by
the low cut-off frequency of RC filters used to reduce external
electromagnetic noise as well as by high resistance of the sample.

\section{EXPERIMENTAL RESULTS}
\label{sect:results}

\begin{figure}
   \begin{center}
   \vspace*{-0.1in}
   \includegraphics[height=8cm,clip=]{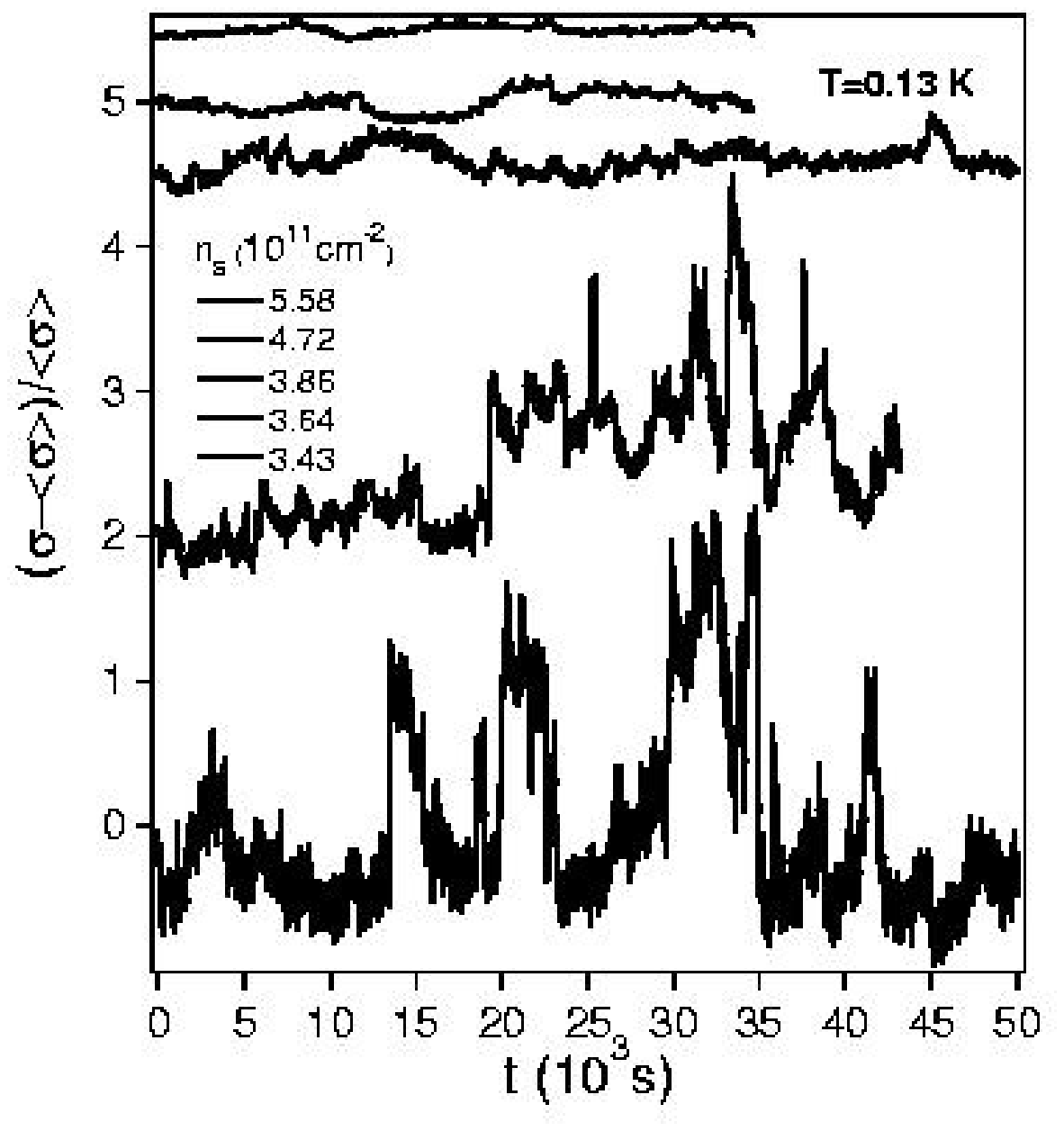}
   \includegraphics[height=8cm,clip=]{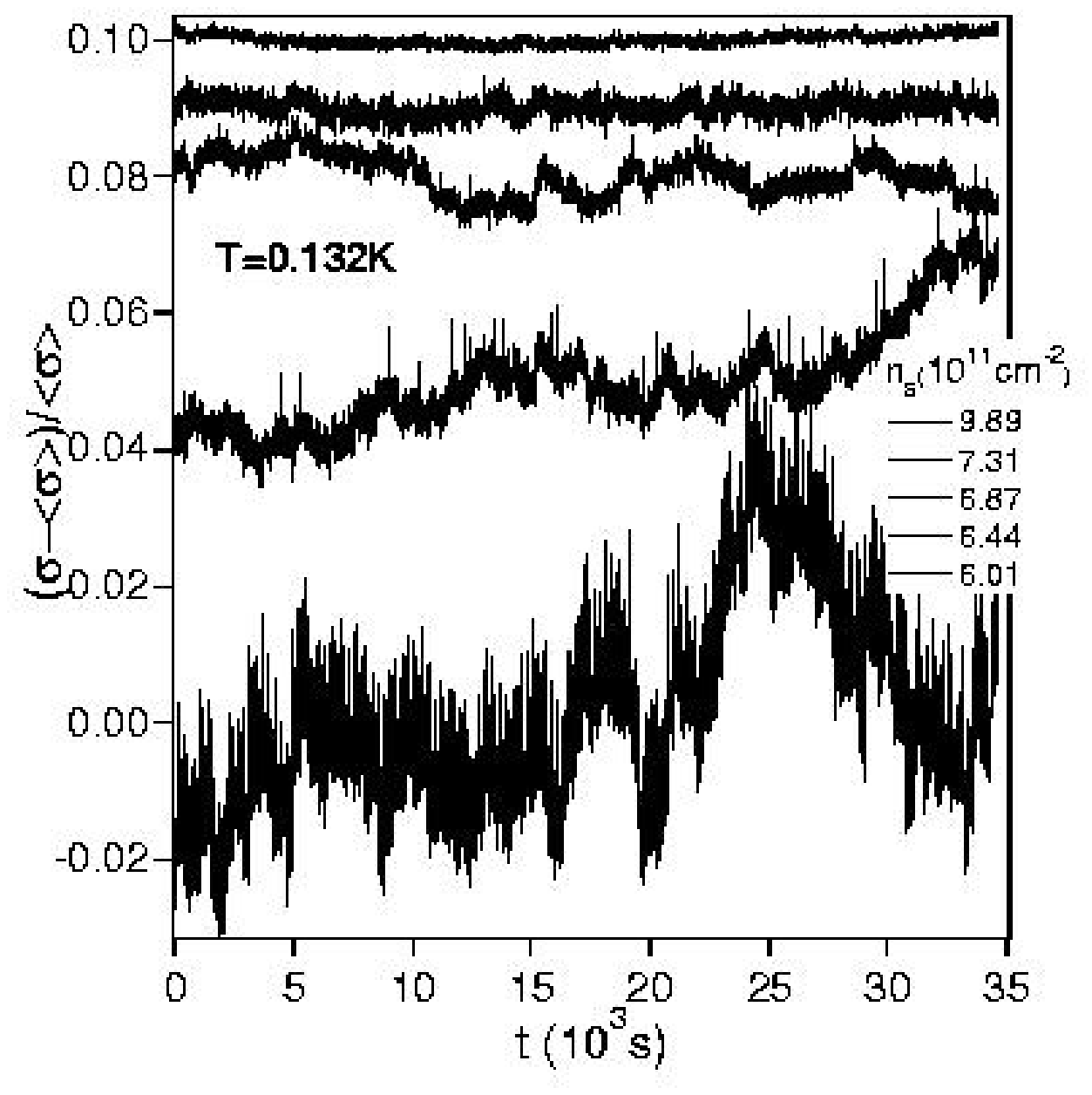}
   \end{center}
   \caption
   { \label{datasb}
Relative fluctuations of $\sigma$ {\it vs.} time for different
$n_s$ in a sample with high disorder at $T=0.13$~K.
$\langle\sigma\rangle$ is the time-averaged conductivity.
Different traces have been shifted for clarity; the lowest $n_s$
is at the bottom and the highest $n_s$ at the top.  The noise
decreases dramatically, from $\sim$100\% to less than 1\%, with
the increasing $n_s$.  The character of the noise also changes
dramatically as $n_s$ is varied.}
   \end{figure}

\begin{figure}
   \begin{center}
   \includegraphics[height=10cm]{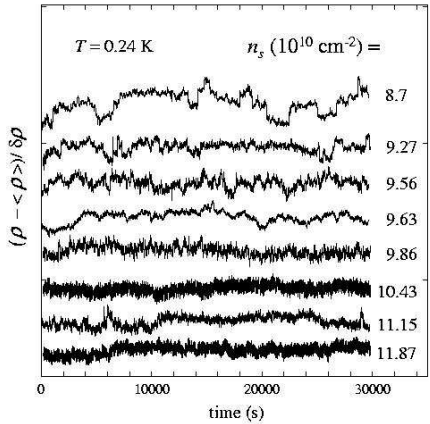}
   \end{center}
   \caption
   { \label{datajj}
Resistance noise in a high-mobility (low disorder) sample for
several $n_s$ shown on the plot.
$(\rho-\langle\rho\rangle)/\delta\rho$ is plotted
($\delta\rho^{2}=$variance, $\rho$ -- resistivity) in order to
make the change in the character of the noise with $n_s$ more
apparent. Different traces have been shifted vertically for
clarity. }
   \end{figure}

In both types of samples, we have observed strong fluctuations of
$\sigma$ with time at low $n_s$ and $T$.  Figure \ref{datasb}
shows the fluctuations of
$(\sigma-\langle\sigma\rangle)/\langle\sigma\rangle$
($\langle\ldots\rangle$ represents averaging over time intervals
of, typically, several hours) in a low-mobility sample
%-------------
%-------------
for several $n_s$ at $T=0.13$~K.  It is quite striking that, for
the lowest $n_s$, the fluctuation amplitude is of the order of
100~\%. In addition to rapid, high-frequency fluctuations, both
abrupt jumps and slow changes over periods of several hours are
also evident.  The amplitude of the fluctuations decreases
dramatically with increasing either $n_s$ (Fig.~\ref{datasb}) or
$T$, as discussed in more detail below. Perhaps even more
interesting is the change in the character of the noise: at high
enough $n_s$, the slow modulations and the discrete events are no
longer apparent in the raw data and, in fact, the variance of the
noise no longer varies with time. Figure \ref{datajj} shows that
similar behavior is observed in the resistance noise of
high-mobility samples. Here we choose to plot
$(\rho-\langle\rho\rangle)/\delta\rho$, where
$\delta\rho=\langle(\rho-\langle\rho\rangle)^{2}\rangle^{1/2}$ and
resistivity $\rho=1/\sigma$, in order to make the changes in the
character of the noise with the variation of $n_s$ more apparent.
In order to try to understand the origin of the observed noise, it
is important to study the transport characteristics first, because
they provide information on the mechanism of conduction in the
regime of interest.
%------------

%%-----------------------------------------------------------
\subsection{Transport}
\label{sect:trans} 
\begin{figure}
   \begin{center}
   \includegraphics[height=9cm]{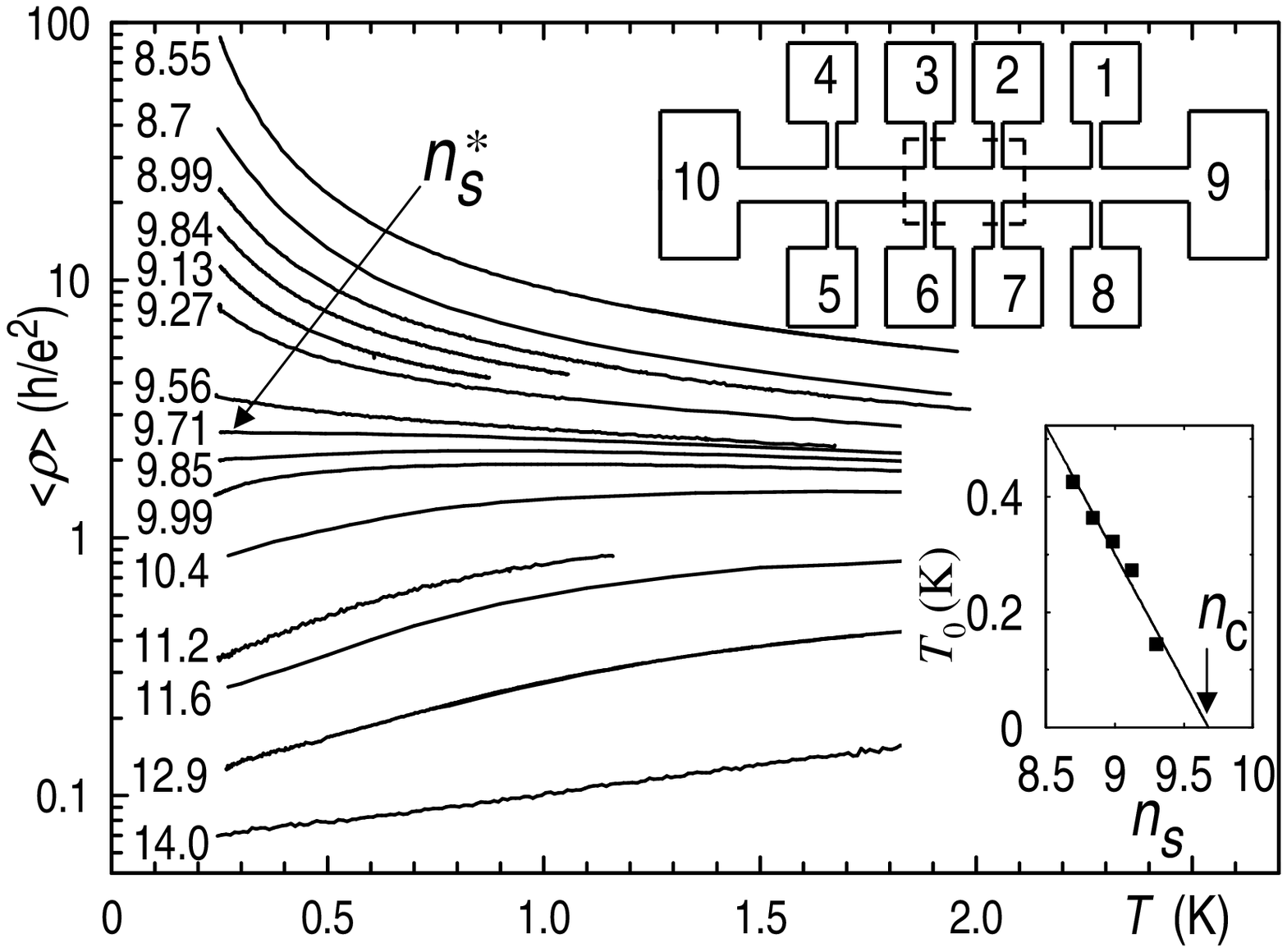}
   \end{center}
   \caption
   { \label{raverage}
High-mobility sample: $\langle\rho\rangle$ {\textit vs.} $T$ for
$n_s(10^{10}$cm$^{-2})=8.55$, 8.70, 8.84, 8.99, 9.13, 9.27, 9.56,
9.71, 9.85, 9.99, 10.4, 11.2, 11.6, 12.9, 14.0 (from the top).
Insets: a schematic of the sample, and activation energies
{\textit vs.} $n_s$; $n_c\approx n_{s}^{\ast}$.}
   \end{figure}

Transport properties of high-mobility samples
almost identical to ours have been studied extensively by several
groups \cite{heemskerksamples1,heemskerksamples2,SAK2001}.
Naturally, we find similar results for the behavior of the
time-averaged resistivity $\langle\rho\rangle$ as a function of
$T$ for different $n_s$, as shown in Fig. \ref{raverage}.
%-------------
%-------------
For the lowest $n_s$ and $T$, the data are described by an
activated form $\langle\rho\rangle\propto\exp (T_0/T)$,
corresponding to transport in the insulating regime. The vanishing
of $T_0$ is often used as a criterion to determine
$n_c$~\cite{Pudalovactivated,Shashkin}, the critical density for
the MIT. Using this method, we find that $n_c\approx9.7\times
10^{10}$cm$^{-2}$ (Fig.~\ref{raverage} inset). In this sample,
$d\langle\rho\rangle/dT$ changes sign at $n_{s}^{\ast}\approx
9.7\times 10^{10}$cm$^{-2}$, so that $n_c\approx n_{s}^{\ast}$ in
agreement with other studies~\cite{Pudalovactivated,Shashkin}.  We
point out, however, that a small but systematic difference of a
few percent has been
reported~\cite{Pudalov_drop,Altshuler_weakloc} such that
$n_{s}^{\ast}>n_c$.  At even higher $n_s$, a pronounced drop of
$\langle\rho\rangle$ with decreasing $T$ is observed. This strong
metallic temperature dependence of resistivity has been a subject
of considerable research effort \cite{SAK2001}, and some progress
has been made recently in understanding its origin
\cite{Zala,Punnoose}.

\begin{figure}
   \begin{center}
   \vspace*{-1.0in}
   \includegraphics[height=16cm]{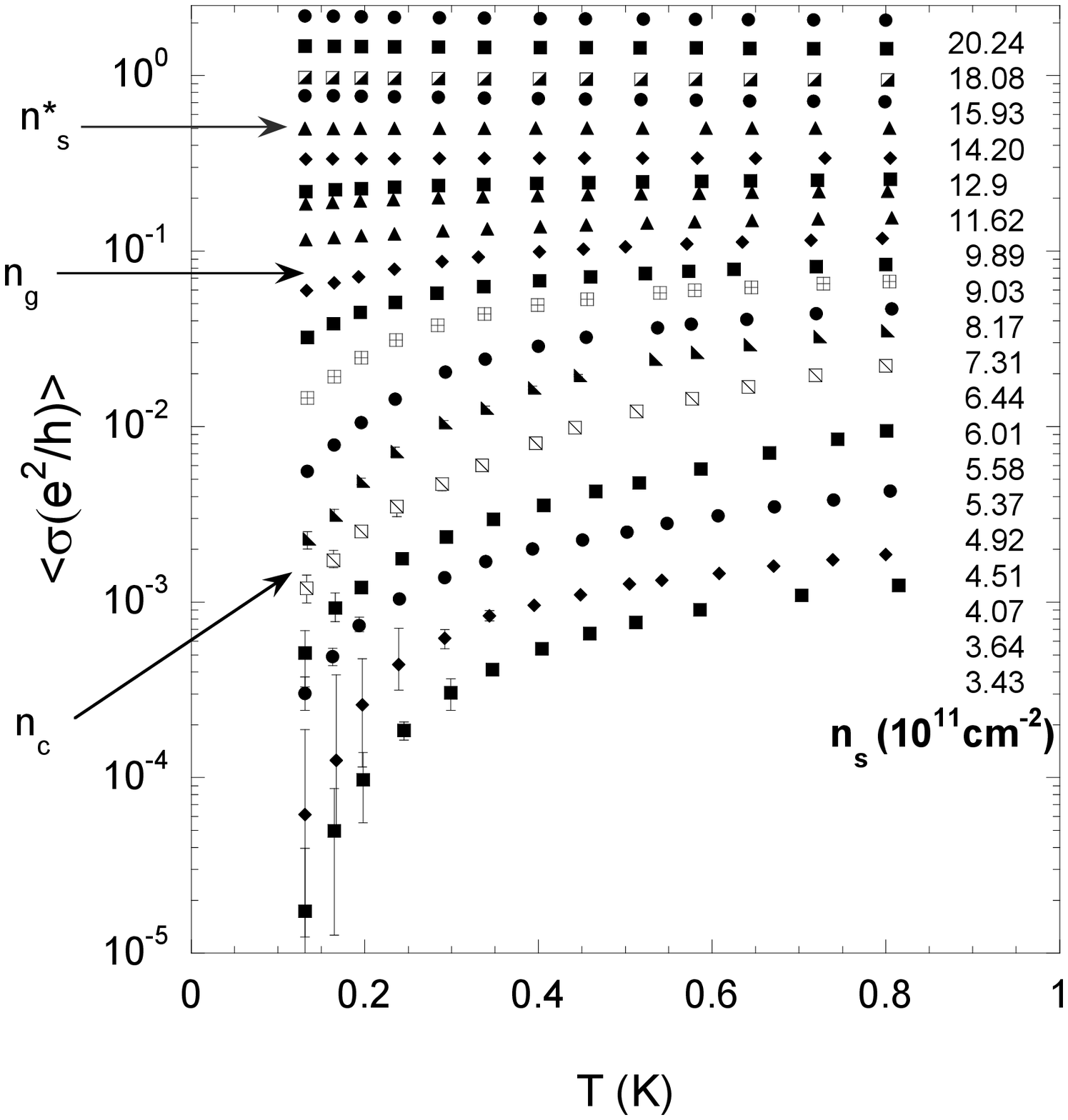}
   \end{center}
   \vspace*{-1.0in}
   \caption
   { \label{saverage}
Low-mobility sample: $\langle\sigma\rangle$ {\it vs.} $T$ for
different $n_s$. The data for many other $n_s$ have been omitted
for clarity.  The error bars show the size of the fluctuations.
$n_{s}^{\ast}$, $n_g$ (glass transition density), and $n_c$ (critical density
for the metal-insulator transiton) are marked by arrows.  They were
determined as explained in the main text.}
   \end{figure}

\begin{figure}
   \begin{center}
   \vspace*{-0.3in}
   \includegraphics[height=19cm]{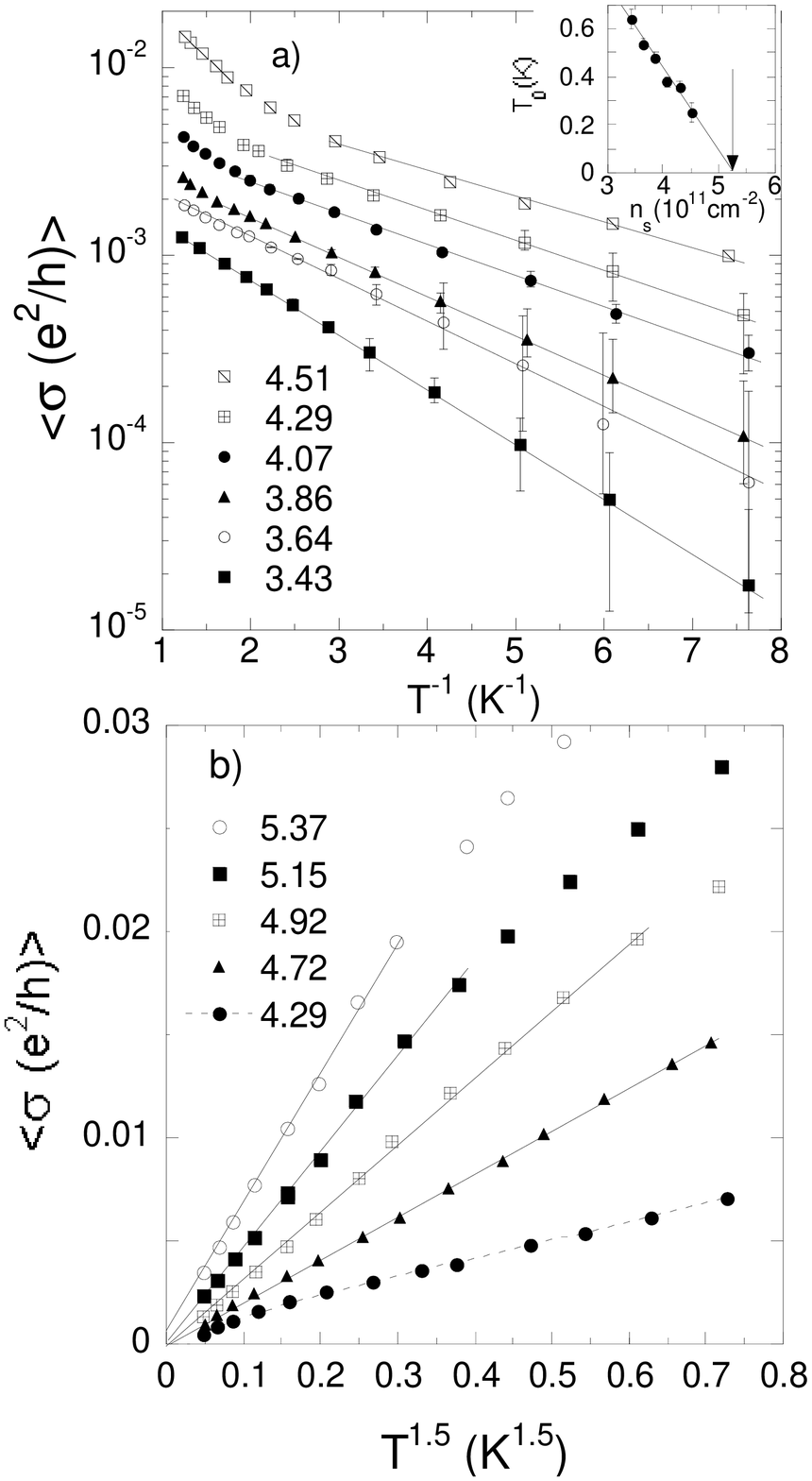}
   \end{center}
   %\vspace*{-0.7in}
   \caption
   { \label{act}
Low-mobility sample: a) $\langle\sigma\rangle$ {\it vs.} $T^{-1}$
for several $n_s (10^{11}$cm$^{-2})$ in the insulating regime. The
error bars show the size of the fluctuations, and the lines are
fits to $\langle\sigma\rangle\propto\exp(-T_0/T)$.  Inset: $T_0$
{\it vs.} $n_s$ with a linear fit, and an arrow showing $n_c$. b)
$\langle\sigma\rangle$ {\it vs.} $T^{1.5}$ for a few $n_s
(10^{11}$cm$^{-2})$ near $n_c$. The solid lines are fits; the
dashed line is a guide to the eye, clearly showing insulating
behavior ($\langle\sigma(T\rightarrow 0)\rangle=0$).}
   \end{figure}

On the other hand, the low-mobility samples used in our work have
not been studied before.  We find that the behavior of
$\langle\sigma (n_s,T)\rangle$ in these samples (Fig. \ref{saverage}) is
qualitatively
similar to that of high-mobility devices. At the highest $n_s$,
for example, low-mobility samples exhibit a metalliclike behavior
with $d\langle\sigma\rangle/dT<0$ (\emph{i. e.}
$d\langle\rho\rangle/dT>0$). The change of $\langle\sigma\rangle$
in a given $T$ range, however, is small (only 6\% for the highest
$n_s=20.2\times10^{11}$cm$^{-2}$) as observed in other Si MOSFETs
with a large amount of disorder~\cite{Pudalov_drop}.
$d\langle\sigma\rangle/dT$ changes sign when
$\langle\sigma(n_{s}^{\ast})\rangle=0.5~e^{2}/h$ similar to other
2D samples~\cite{SAK2001}.
%-------------
%-------------
For the lowest $n_s$, the data are again best described by the
simply activated form $\langle\sigma\rangle\propto\exp (-T_{0}/T)$
[Fig.~\ref{act}(a)].  $T_0$ decreases linearly with increasing
$n_s$ (Fig.~\ref{act}(a) inset), and vanishes at
$n_c\approx5.2\times10^{11}$cm$^{-2}$.
%-------------
%-------------
Surprisingly, we find that, close to $n_c$, the data are best
described by the metallic power-law behavior
$\langle\sigma(n_s,T)\rangle=a(n_s)+b(n_s)T^{x}$ with
$x\approx1.5$ [Fig.~\ref{act}(b)].  The fitting parameter $a(n_s)$
is relatively small and, in fact, vanishes for $n_s
(10^{11}$cm$^{-2})=4.72$ and 4.92.  Such a simple power-law
$T$-dependence of $\sigma$, given by $\langle\sigma
(n_c,T)\rangle\propto T^x$, is consistent with the one expected in
the quantum critical region (QCR) of the MIT based on general
arguments~\cite{Belitz}, and with the behavior observed in 3D
systems~\cite{Belitz} and other Si MOSFETs~\cite{Feng-novel}
within the QCR.  Moreover, the $T^{3/2}$ correction is consistent
with the recent theory for the metallic glass phase \cite{Denis}.
Therefore, based on the analysis of $\langle\sigma (n_s,T)\rangle$
in both insulating regime and QCR, we conclude that the critical
density $n_c=(5.0\pm 0.3)\times 10^{11}$cm$^{-2}$, which is more
than a factor of two smaller than $n_{s}^{\ast}$. Such a large
difference between $n_c$ and $n_{s}^{\ast}$ is attributed to a
much higher amount of disorder in these samples than in
high-mobility Si
MOSFETs~\cite{Pudalov_drop,Altshuler_weakloc,Pudalovactivated,Shashkin,JJPRL02}.

%%-----------------------------------------------------------
\subsection{Noise}
\label{sect:noi} We begin our discussion of noise by considering
low-mobility samples first.  A simple analysis shows
\cite{SBPRL02} that $\delta\sigma/\langle\sigma\rangle$
($\delta\sigma^2=$ variance) does not depend on $n_s$ and $T$ at
high enough $n_s$. However, below a certain density $n_g = (7.5\pm
0.3)\times10^{11}$cm$^{-2}$, which does not seem to depend on $T$,
an enormous increase of $\delta\sigma/\langle\sigma\rangle$ is
observed with decreasing either $n_s$ or $T$.  It is interesting
that $\delta\sigma/\langle\sigma\rangle$ does not exhibit any
special features near $n_c$ or $n_{s}^{\ast}$.

\begin{figure}
   \begin{center}
   \includegraphics[height=8cm,clip=]{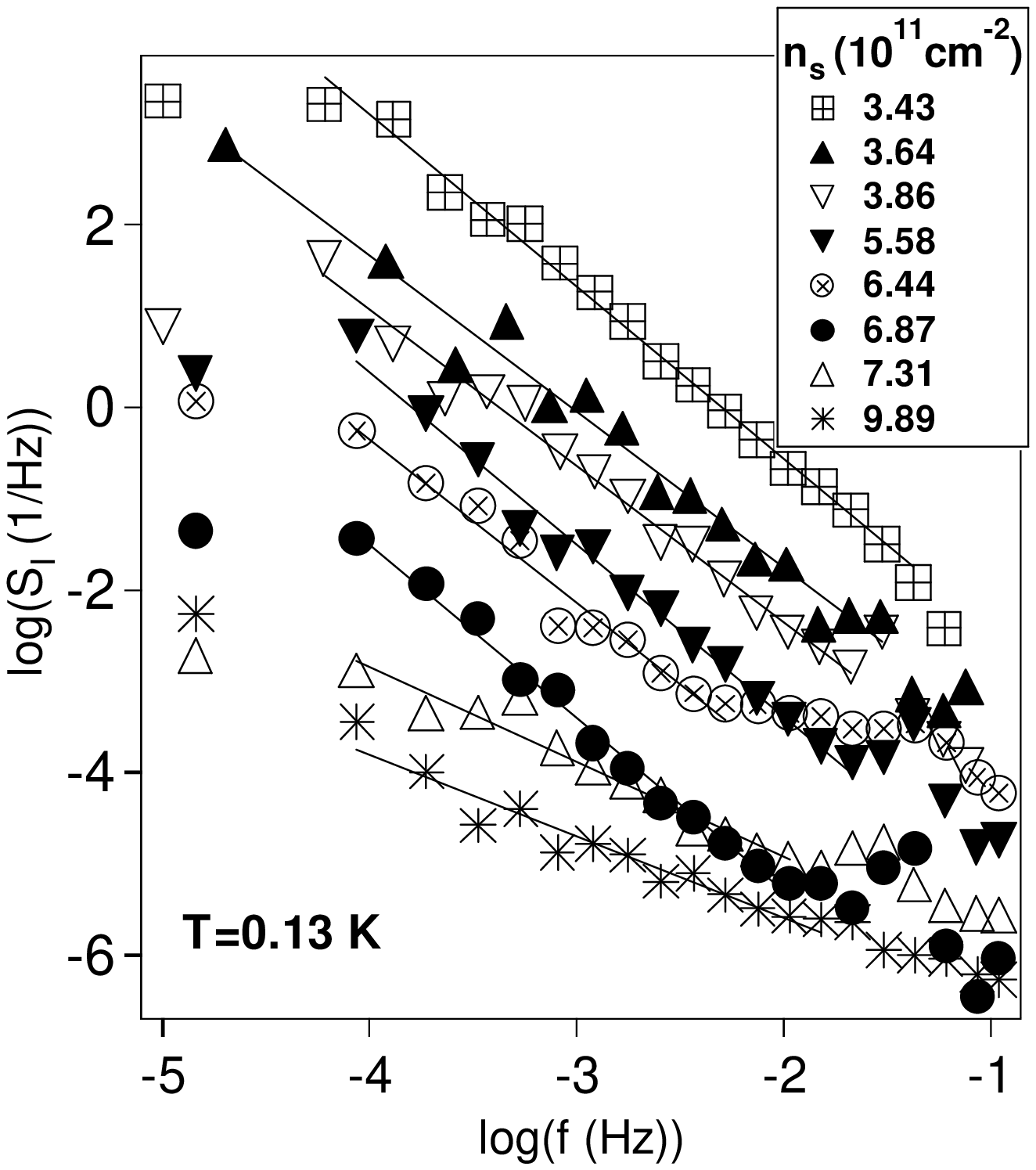}
   \end{center}
   \caption
   { \label{sf}
   The averaged noise power spectra $S_I\propto 1/f^{\alpha}$ {\it
vs.} $f$ for several $n_s$ in a low-mobility sample. Solid lines
are linear least-squares fits with the slopes equal to $\alpha$.}
   \end{figure}

\begin{figure}[h]
   \begin{center}
   \vspace*{-0.6in}
   \includegraphics[height=12cm,clip=]{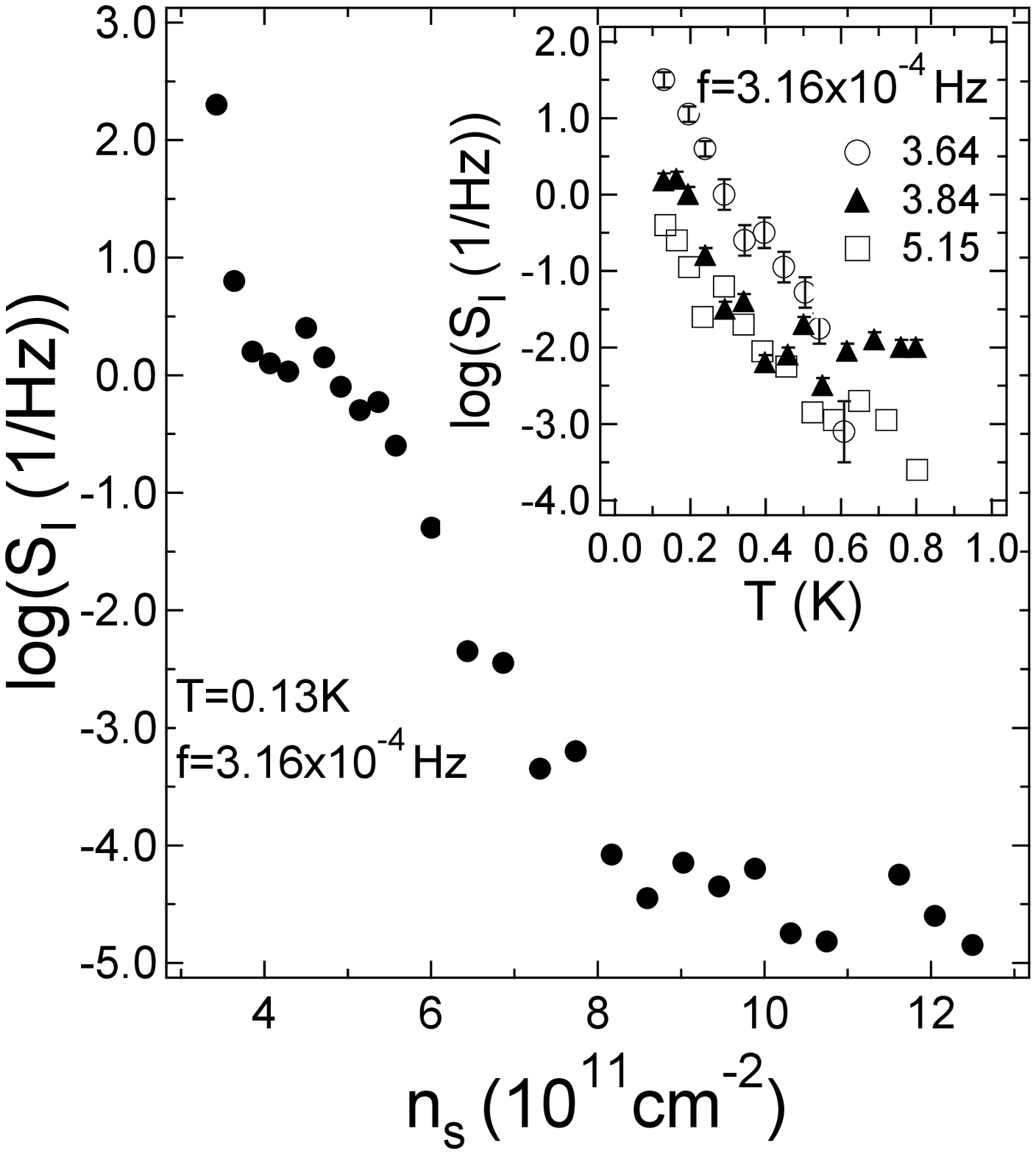}
   \end{center}
   \vspace*{-1.0in}
   \caption
   { \label{sn}
   The normalized noise power $S_I(f=3.16\times 10^{-4}$Hz) {\it
vs.} $n_s$ in a low-mobility sample at $T=0.13$~K.  Below
$n_g\approx 7.5\times 10^{11}$cm$^{-2}$, the noise increases
exponentially with decreasing $n_s$.  Inset: $S_I$ \textit{vs.}
$T$ for three different $n_s(10^{11}$cm$^{-2})$ given on the
plot.}
   \end{figure}

A more detailed study of the noise has been carried out by
calculating the normalized power spectra $S_{I}(f)=S(I,f)/I^{2}$
($f$--frequency) of
$(\sigma-\langle\sigma\rangle)/\langle\sigma\rangle$ for all $n_s$
and $T$. Most of the spectra were obtained in the
$f=(10^{-4}-10^{-1})$~Hz bandwidth, where they follow the
well-known empirical law $S_I\propto
1/f^{\alpha}$~\cite{Hooge,Weiss88}.  The background noise was
measured by setting $I=0$ for all $n_s$ and $T$.  It was always
white and usually several orders of magnitude smaller than the
sample noise.  Nevertheless, a subtraction of the background
spectra was always performed, and the power spectra of the device
noise were averaged over frequency bands ($\leq$ an octave). Some
of the resulting $S_I$ are presented in Fig.~\ref{sf}.
%-------------
%-------------
At the highest $n_s$, $S_I(f)$ does not depend on $n_s$. However,
it is obvious that, by reducing $n_s$ below $n_g$, $S_I$ increases
enormously, by up to six orders of magnitude at low $f$, as shown
in Fig. \ref{sn}.
%-------------
%-------------
This striking increase of the slow dynamic contribution to the
conductivity reflects a {\em sudden and dramatic slowing down of
the electron dynamics}.  This is attributed to the freezing of the
electron glass.

We also find that, for $n_s<n_g$, $S_I(f)$ {\em increases}
exponentially with decreasing $T$ (Fig.~\ref{sn} inset),
consistent with early studies on Si MOSFETs where $dS_R/dT<0$ was
observed for $T=1.5$, 4.2~K \cite{Adkins}. Such a temperature
dependence shows that the noise in our system cannot be explained
by the models of thermally activated charge
trapping~\cite{Weiss88,Dutta,Rogers84}, noise generated by
fluctuations of $T$~\cite{VC}, and a model of noise near the
Anderson transition~\cite{Cohen1,Cohen2}.  Likewise, the models of
noise in the Mott and Efros-Shklovskii variable-range hopping
(VRH) regimes
\cite{Shklovskii-hopnoi1,Shklovskii-hopnoi2,Kozub,Shtengel}
predict either $dS_R/dT>0$ or a saturation of $S_R$ below
10-100~Hz, both in clear disagreement with the experiment.
Therefore, the observed noise cannot be a result of single
electron hops even when Coulomb interactions are included through
the Coulomb gap.  In principle, it is possible that the VRH models
of noise
\cite{Shklovskii-hopnoi1,Shklovskii-hopnoi2,Kozub,Shtengel} may
not be applicable to our low-mobility samples with their
relatively short and wide geometry.  In such cases, one may expect
\cite{RR,GM} that the percolation cluster will be reduced to a
small number of parallel hopping chains, resulting in a weaker
temperature dependence of conductivity. Measurements
\cite{Pokrovskii} of $1/f$ noise in 2D hopping of electrons in
GaAs samples with lengths (0.5 and 1~$\mu$m) comparable to ours
and performed at higher $T$ ($> 4.2$~K) and $f$ (1 Hz), have shown
a power-law increase of noise with decreasing $T$, in agreement
with the VRH models of noise \cite{Kozub}.  In the shortest
sample, with the length $\simeq0.2 \mu$m, the cluster was reduced
down to a set of well separated linear chains \cite{Laiko,Orlov}.
In that sample, the temperature dependence of noise was found
\cite{Pokrovskii} to be even \emph{weaker} than in longer samples,
which is exactly the opposite of what is observed in our samples.
These results show clearly that the noise in our low-mobility
samples cannot be explained based on the model of hopping chains.
We also note that a small number of parallel hopping chains would
result in large ($\sim 100$\%) fluctuations of conductivity with
the gate voltage, which moves the Fermi energy through different
sets of localized states. While such fluctuations have been
observed in other experiments \cite{Laiko,Orlov,DP90,DP91}, here
we find that the fluctuations of $\langle\sigma\rangle$ with $V_g$
are only of the order 0.1\% in the range of $n_s$ where noise has
been studied. On the other hand, an increase of noise at low $T$
similar to our results has been observed in mesoscopic spin
glasses~\cite{Israeloff89,jjprl98,Neutt}, in wires in the quantum
Hall regime for tunneling through localized states~\cite{Wrobel},
and in Si quantum dots in the Coulomb blockade
regime~\cite{Molenkamp}.

\begin{figure}
   \begin{center}
   \includegraphics[height=8cm,clip=]{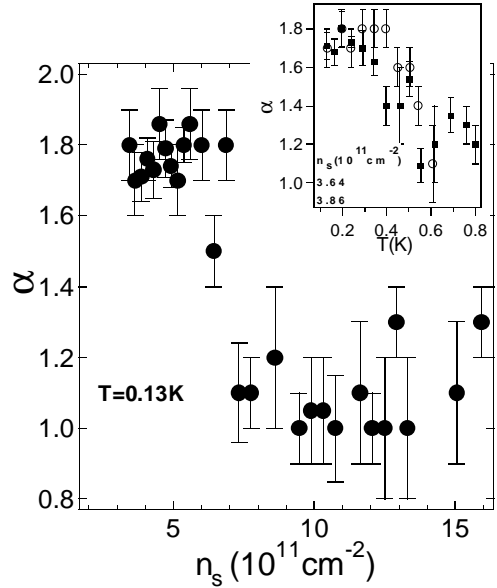}
   \end{center}
   \vspace*{-0.3in}
   \caption
   { \label{expalpha}
Low-mobility sample: at $n_s\approx n_g$, the exponent $\alpha$
exhibits a sharp jump from $\approx 1$ at high $n_s$ (``pure''
$1/f$ noise) to $\approx1.8$ at low $n_s$.  Inset: $\alpha$
\textit{vs.} $T$ for two different $n_s(10^{11}$cm$^{-2})$ (3.64
-- open symbols, 3.86 -- solid symbols) in the glassy phase. }
   \end{figure}

Perhaps the most striking feature of our data is the sharp jump of
the exponent $\alpha$ at $n_s\approx n_g$ (Fig. \ref{expalpha}).
While $\alpha\approx1$ (``pure'' $1/f$ noise) for $n_s>n_g$,
$\alpha\approx 1.8$ below $n_g$, reflecting a sudden shift of the
spectral weight towards lower frequencies.  This is yet another
manifestation of the sudden and dramatic slowing down of the
electron dynamics at $n_g$.  Similar large values of $\alpha$ have
been observed in some spin glasses above the
MIT~\cite{jjprl98,Neutt}, and in submicron wires~\cite{Wrobel}.
This will be discussed in more detail below.  Here we point out
that the onset of glassy dynamics occurs on the metallic side of
the MIT, at the density $n_g>n_c$ (where VRH models are clearly not
applicable!). This implies the existence of
the metallic glass phase for $n_c<n_s<n_g$, which is actually
consistent with recent predictions of the model of interacting
electrons near a disorder-driven MIT \cite{Darko}.  We also note
that, in the glassy phase, $\alpha$ decreases with increasing $T$
(Fig.~\ref{expalpha} inset).  As a result of such $\alpha(T)$: (i)
the large values of
%-------------
%-------------
$\alpha$ are observable only at relatively low $T$, and (ii) the
rise in $\alpha$ with decreasing $n_s$ becomes smoother,
\textit{i. e.} less sharp at higher $T$.

\begin{figure}
   \begin{center}
   %\vspace*{-1.0in}
   \includegraphics[height=7cm,clip=]{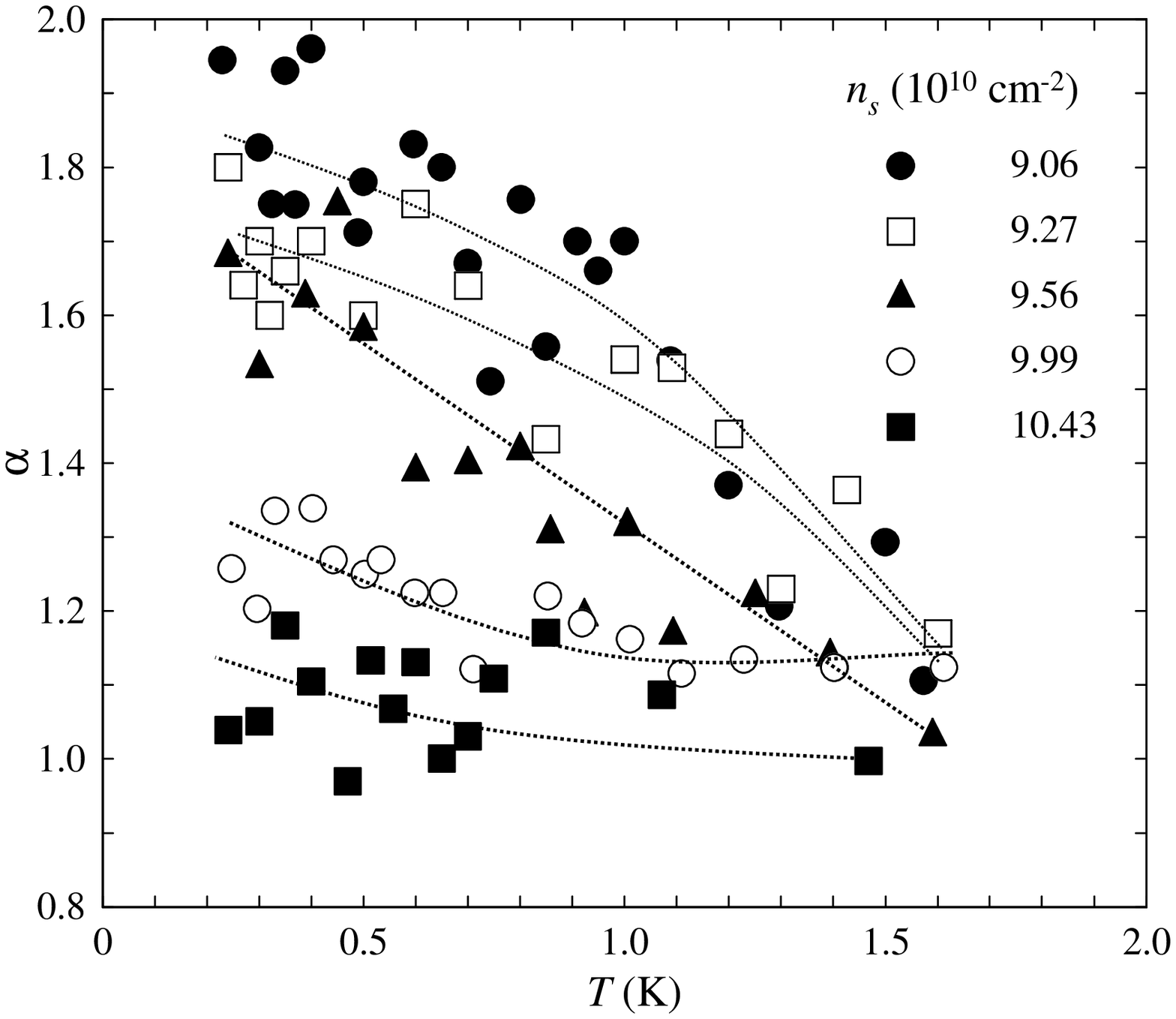}
   \end{center}
   \vspace*{-0.2in}
   \caption
   { \label{jjexpalpha}
$\alpha$ \textit{vs.} $T$ for several $n_s$ in a high-mobility
sample; $n_g\approx 10\times 10^{10}$cm$^{-2}$.  Lines are guides
to the eye.  }
   \end{figure}

In high-mobility samples, the time series of the relative changes
in resistance $\Delta R(t)/\langle R\rangle$ and the corresponding
power spectra $S_R(f)$ have been studied in detail, and
qualitatively the same behavior has been found \cite{JJPRL02}. At
high $n_s$, the low-frequency \textit{(e. g.} 1 mHz) noise power
depends rather weakly on both $T$ and $n_s$.  In the vicinity of
$n_g\approx 10\times 10^{10}$cm$^{-2}$, however, a dramatic change
in the behavior of $S_R$ is observed at low $T$.  The noise
amplitude starts to increase strongly with decreasing $n_s$, and
the exponent $\alpha$ jumps from $\approx1$ to $\approx1.8$, which
is again attributed to the freezing of the electron glass.  The
temperature dependence of $\alpha$ for these samples is shown in
Fig.~\ref{jjexpalpha}. We point out that these samples are much
longer than the low-mobility ones and, in fact, have a completely
opposite geometry (long and narrow, as opposed to short and wide).
This provides further evidence that the observed behavior of noise
cannot be attributed to geometric effects.
%-------------
%-------------

Since the amount of disorder in these samples is considerably
lower than in the low-mobility ones, the absolute value of $n_g$
is, not surprisingly, almost an order of magnitude lower. However,
in addition to affecting the values of $n_c$, $n_g$, and
$n_{s}^{\ast}$, the disorder plays another, nontrivial role.  In
particular, $n_c$ and $n_g$ were found to differ from each other
considerably in low-mobility devices ($n_c$, $n_g$, and
$n_{s}^{\ast}$ were 5.0, 7.5, and 12.9, respectively, in units of
$10^{11}$cm$^{-2}$), whereas in high-mobility devices $n_g$ is at
most a few percent higher than $n_c$. Therefore, the emergence of
glassy dynamics here seems almost to coincide with the MIT.
Obviously, the size of the metallic ($n_c<n_s<n_g$) glass phase
depends strongly on disorder, in agreement with theoretical
predictions~\cite{Darko}.

We have established that the exponent $\alpha\approx 1$ in the 2D
metallic phase (above $n_g$) in both low- and high-mobility
samples.  On the other hand, $\alpha\approx 1.8$ in the glassy
phase.  In general, such high values of $\alpha$ may be also
obtained if noise results from a superposition of a small number
of independent two-state systems (TSS)~\cite{Dutta,Weiss88}.
However, even though some distinguishable discrete events can be
seen at low $n_s$ (Figs. \ref{datasb} and \ref{datajj}), they do
not show the characteristic repetitive form of stable TSS.  On the
contrary, both the shape and the magnitude of noise exhibit
random, nonmonotonic (which exclude aging) changes with time.  A
quantitative measure of such spectral wandering is the so-called
second spectrum $S_2(f_2,f)$, which is the power spectrum of the
fluctuations of $S_{R}(f)$ with time~\cite{Weiss93,WeissMMM},
{\textit i.~e.} the Fourier transform of the autocorrelation
function of the time series of $S_{R}(f)$. If the fluctuators
({\textit e.~g.} TSS) are not correlated, $S_2(f_2,f)$ is white
(independent of $f_2$)~\cite{WeissMMM,Weiss88,Weiss93} and equal
to the square of the first spectrum.  Such noise is called
Gaussian. On the other hand, $S_2$ has a nonwhite character,
$S_2\propto 1/f_{2}^{1-\beta}$, for interacting
fluctuators~\cite{WeissMMM,Weiss88,Weiss93}. Therefore, the
deviations from Gaussianity provide a direct probe of correlations
between fluctuators.

\begin{figure}
   \begin{center}
   \vspace*{0.2in}
   \includegraphics[height=6cm,clip=]{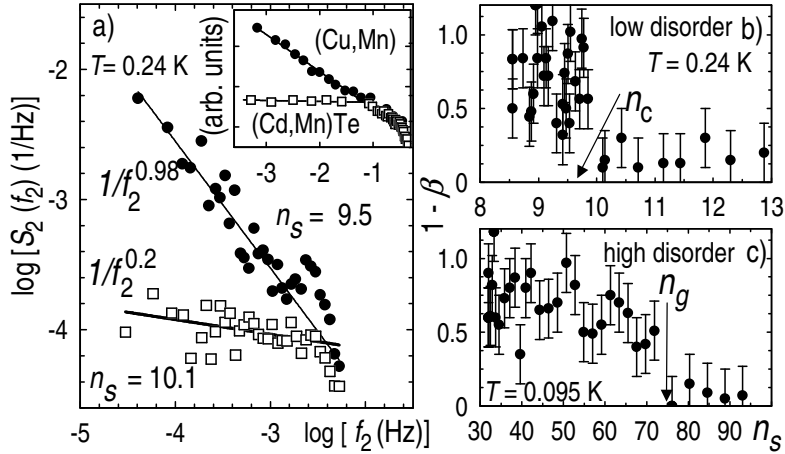}
   \end{center}
   %\vspace*{-0.8in}
   \caption
   { \label{second}
(a) Second spectral density $S_2(f_2)$ {\textit vs.} $f_2$ for
$n_s (10^{10}$cm$^{-2})$ shown on the plot; $f_L=1$~mHz.  Inset:
$S_2(f_2)$ for spin glasses Cu$_{0.91}$Mn$_{0.09}$
\cite{WeissMMM,Weiss93} and Cd$_{0.93}$Mn$_{0.07}$Te
\cite{jjprl98}. Solid lines are fits. Exponent $1-\beta$ {\textit
vs.} $n_s$ for (b) high-mobility and (c) low-mobility samples. The
low-mobility device is the same as the one studied in
Ref.~\cite{SBPRL02}.}
   \end{figure}

We investigate $S_2$ using digital filtering~\cite{Seid96a,Abke}
in a given frequency range $f=(f_L,f_H)$ (usually $f_H=2f_L$). The
normalized second spectra, with the Gaussian background
subtracted, are shown in Fig.~\ref{second}(a) for two $n_s$, just
above and just below $n_g$.
%-------------
%-------------
It is clear that there is a striking difference in the character
of the two spectra.  Similar differences are observed between
various spin glasses (Fig.~\ref{second}(a) inset), where $S_2$ is
white~\cite{jjprl98} in the absence of long range interactions,
and nonwhite~\cite{WeissMMM,Weiss93} when long range RKKY
interaction leads to hierarchical glassy dynamics~\cite{Ogielski}.
A detailed dependence of the exponent $(1-\beta)$ on $n_s$ has
been determined for both high- and low-mobility samples
(Figs.~\ref{second}(b) and (c), respectively). In both cases,
$S_2$ is white for $n_s>n_g$, indicating that the observed $1/f$
noise results from uncorrelated fluctuators.  It is quite
remarkable that $S_2$ changes its character in a dramatic way
exactly at $n_g$ in both types of samples.  For $n_s<n_g$, $S_2$
is strongly non-Gaussian, which demonstrates that the fluctuators
are strongly correlated.  This, of course, rules out independent
TSS (such as traps) as possible sources of noise when $n_s<n_g$.
In fact, a sudden change in the nature of the fluctuators
({\textit i.~e.} correlated {\textit vs.} uncorrelated) as a
function of $n_s$ rules out {\em any} traps, defects, or a highly
unlikely scenario that the observed glassiness may be due to some
other time dependent changes of the disorder potential itself.
Instead, it provides an unambiguous evidence for the onset of
glassy dynamics in a 2D electron system at $n_g$.

\begin{figure}
   \begin{center}
%   \vspace*{-0.2in}
   \includegraphics[height=6cm,clip=]{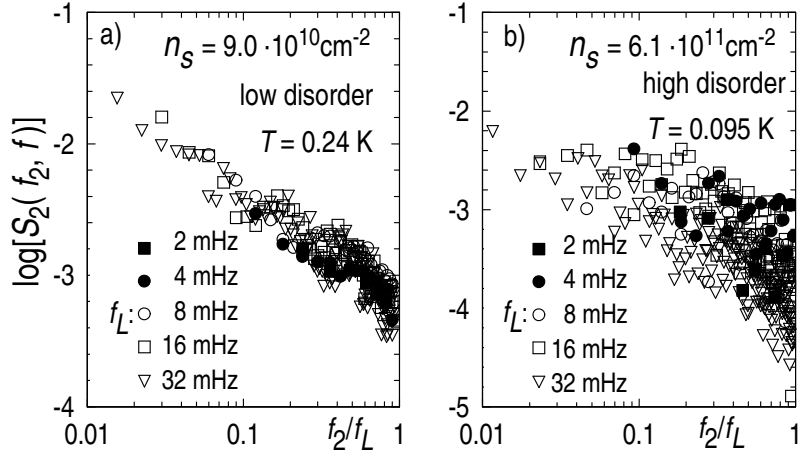}
   \end{center}
   %\vspace*{-0.8in}
   \caption
   { \label{scaling}
Scaling of $S_2$ measured at frequency octaves $f=(f_L,2f_L)$ for
(a) high-mobility and (b) low-mobility samples.}
   \end{figure}

In the studies of spin glasses, the scaling of $S_2$ with respect
to $f$ and $f_2$ has been used~\cite{WeissMMM,Weiss93} to unravel
the glassy dynamics and, in particular, to distinguish generalized
models of interacting droplets or clusters ({\textit i.~e.} TSS)
from hierarchical pictures. In the former case, the low-$f$
noise comes from a smaller number of large elements because they
are slower, while the higher-$f$ noise comes from a larger
number of smaller elements, which are faster.  In this picture,
which assumes compact droplets and short-range interactions between
them, big elements are more likely to interact than small ones and,
hence, non-Gaussian effects
and $S_2$ will be stronger for lower $f$.  $S_2(f_2,f)$, however,
need to be compared for fixed $f_2/f$, {\textit i.~e.} on time
scales determined by the time scales of the fluctuations being
measured, since spectra taken over a fixed time interval average
the high-frequency data more than the low-frequency data.
Therefore, in the interacting ``droplet'' model, $S_2(f_2,f)$
should be a decreasing function of $f$ at constant $f_2/f$.  In
the hierarchical picture, on the other hand, $S_2(f_2,f)$ should
be scale invariant: it should depend only on $f_2/f$, not on the
scale $f$~\cite{WeissMMM,Weiss93}. Fig.~\ref{scaling} shows that
no systematic dependence of $S_2$ on $f$ is seen in our samples.
The observed scale invariance of $S_2(f_2,f)$
%-------------
%-------------
signals that the system wanders collectively between many
metastable states related by a kinetic hierarchy.  Metastable
states correspond to the local minima or ``valleys'' in the free
energy landscape, separated by barriers with a wide, hierarchical
distribution of heights and, thus, relaxation times. Intervalley
transitions, which are reconfigurations of a large number of
electrons, thus lead to the observed strong, correlated,
$1/f$-type noise, remarkably similar to what was observed in spin
glasses with a long-range correlation of spin
configuration~\cite{WeissMMM,Weiss93}.  We note that, unlike
droplet models~\cite{Fisher1,Fisher2}, hierarchical pictures of
glassy dynamics~\cite{Binder} do allow for the existence of a
finite $T$ (or finite Fermi energy) glass transition in presence
of a symmetry-breaking field, such as the random potential in an
electron glass.

%%-----------------------------------------------------------
\subsection{Slow relaxations and history dependence}
\label{sect:hist} 
\begin{figure}
   \begin{center}
   \vspace*{-0.2in}
   \includegraphics[height=10cm,clip=]{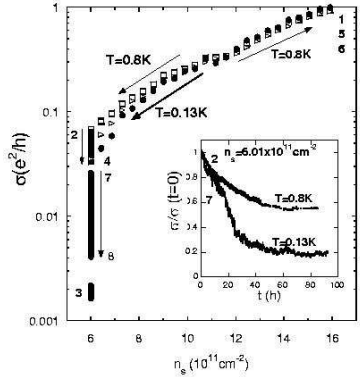}
   \end{center}
   \vspace*{-0.2in}
   \caption
   { \label{hist}
Two different cooling protocols. I $n_s(10^{11}$cm$^{-2})$ was
changed slowly ($\approx 4.5$ hours) from 15.93 to 6.01 at
$T=0.8$~K ($1\rightarrow2$, open squares), and the system was allowed to relax
(top trace in the inset). After $\sigma$ reached a stationary
value, the sample was cooled down to 0.13~K (point 3 on the plot),
where no further relaxation occurred.  The system was then warmed
up (4), and $n_s$ returned to its initial value ($4\rightarrow5$, open
triangles).  This was
followed by the second protocol: II The system was first cooled
down to 0.13~K (point 6, solid circle), and $n_s$ was then changed to
$6.01\times10^{11}$cm$^{-2}$ ($6\rightarrow7$, solid circles) at about the
same rate as
before.  The system was allowed to relax ($7\rightarrow8$, and
bottom trace in the inset).  $\sigma$ reached a stationary value
(point 8) that was, by a factor of 2, different from the value
(point 3) obtained in the first protocol for the same $n_s$ and
$T$, clearly demonstrating history-dependent behavior.}
   \end{figure}

The 2D electron glass in Si MOSFETs also
exhibits other phenomena characteristic of glassy systems. In
particular, for $n_s<n_g$, we have observed history dependent
behavior, and long relaxation times following a large change in
$V_g$.  While further careful investigation is required in order
to study these effects in detail, here we present an example of
the results obtained following two different cooling protocols in
a low-mobility sample.  In the first one, $n_s(10^{11}$cm$^{-2})$
was changed slowly ($\approx4.5$ hours) from 15.93 ($>n_g$) to
6.01 ($<n_g$) at $T=0.8$~K (from point 1 to point 2 in Fig.
\ref{hist}), and then the system was allowed to relax (top trace
%-------------
%-------------
in Fig. \ref{hist} inset).  The conductivity reached a stationary
value after about 70 hours.  The system was then cooled to
$T=0.13$~K (point 3 in Fig. \ref{hist}), where no further
relaxation (\textit{i. e.} monotonic decrease) of $\sigma$ was
observed (the height of ``point'' 3 in Fig. \ref{hist} reflects
fluctuations of $\sigma$ with time).  The system was then warmed
back up to $T=0.8$~K (point 4 in Fig. \ref{hist}), and $n_s$
changed back to its starting value (point 5 in Fig. \ref{hist},
identical to 1).  In the next, second cooling protocol, the system
was first cooled down to 0.13~K (point 6 in Fig. \ref{hist}), and
$n_s(10^{11}$cm$^{-2})$ was then changed from 15.93 to 6.01 (point
7 in Fig. \ref{hist}) at about the same rate as before.  This
time, the relaxation ($7\rightarrow8$ in Fig. \ref{hist}, and
bottom trace in the inset) was nonmonotonic and it took about
80-90 hours for $\sigma$ to reach a stationary value (point 8 in
Fig. \ref{hist}). Most importantly, the final values of $\sigma$
(points 3 and 8) obtained for the same $n_s$ and $T$ following two
different cooling protocols differ by a factor of 2, clearly
demonstrating history-dependent, \textit{i. e.} nonergodic
behavior.

The history dependence of $\sigma$ at low $n_s$ and $T$ was
actually the first ``unusual'' property that we observed in this
system.  In order to study its transport properties and obtain
reproducible values of $\langle\sigma(n_s,T)\rangle$ shown in
Figs. \ref{raverage} and \ref{saverage}, it was necessary to vary
$n_s$ at a relatively high $T$ (we used 0.8~K and $\approx2$~K for
low- and high-mobility samples, respectively).

%%%%%%%%%%%%%%%%%%%%%%%%%%%%%%%%%%%%%%%%%%%%%%%%%%%%%%%%%%%%%
\section{CONCLUSIONS} \label{sect:sections}

We have presented evidence for the freezing of the electron glass
at a well-defined density $n_g$ in a 2DES in Si.  By studying the
statistics of low-frequency resistance noise in Si MOSFETs with a
wide range of characteristics, including a vast difference in the
amount of disorder, we have established that the glassy ordering
of a 2DES near the metal-insulator transition is a universal
phenomenon in Si inversion layers.  Such ordering is observable
only at sufficiently low $T$, and becomes more pronounced with
decreasing $T$.  Glassy freezing occurs in the regime of very low
conductivities but on the metallic side of the MIT.  The size of
the metallic glass phase, which separates the 2D metal and the
(glassy) insulator, depends strongly on disorder, becoming
extremely small in high-mobility samples.  The existence of such a
metallic glass phase and its dependence on disorder are consistent
with theoretical predictions \cite{Darko}.  The glass transition
is manifested by a sudden and dramatic slowing down of the
electron dynamics and by an abrupt change to the sort of
statistics characteristic of complicated multistate systems,
consistent with the hierarchical picture of glassy dynamics and
very similar to spin glasses with long-range correlations. Most
recent studies of noise in parallel magnetic fields have shown
\cite{noiseB1} that the glass transition persists even in fields
such that the 2DES is fully spin polarized.  Therefore, our
results provide strong support to the theoretical proposals that
attempt to describe the 2D metal-insulator transition as the
melting of a Coulomb glass \cite{Pastor,Denis,Darko}.

%%%%%%%%%%%%%%%%%%%%%%%%%%%%%%%%%%%%%%%%%%%%%%%%%%%%%%%%%%%%%
%%  Use following command to specify that graphics file is in
%%  a directory other than this LaTeX source file.
%%  Note use of / to separate subdirectories, for UNIX and Windows OS.
%%\graphicspath{{H:/HANSON/SPIESTY/}}
%%%%%%%%%%%%%%%%%%%%%%%%%%%%%%%%%%%%%%%%%%%%%%%%%%%%%%%%%%%%%
\acknowledgments     %>>>> equivalent to \section*{ACKNOWLEDGMENTS}

We are grateful to the Silicon Facility at IBM, Yorktown Heights
for fabricating low-mobility samples, and to V. Dobrosavljevi\'c
for useful discussions.  This work was supported by NSF Grant No.
DMR-0071668 and NHMFL through NSF Cooperative Agreement No.
DMR-0084173.
%%%%%%%%%%%%%%%%%%%%%%%%%%%%%%%%%%%%%%%%%%%%%%%%%%%%%%%%%%%%%
%%%%% References %%%%%

\bibliography{popovic}   %>>>> bibliography data in report.bib
\bibliographystyle{spiebib}   %>>>> makes bibtex use spiebib.bst

\end{document}